\documentclass[
    ,final            
  ]
  {aipproc}

\layoutstyle{6x9}

\newcommand{\cond}{\langle \bar q q \rangle}
\newcommand{\Od}{{\cal O}}
\newcommand{\im}{\mbox{Im}\,}
\newcommand{\re}{\mbox{Re}\,}

\begin{document}

\title{Chiral symmetry and mesons in hot and dense matter: recent developments}


\author{A.G\'omez Nicola}{
  address={Departamento de F\'{\i}sica
Te\'orica II. Univ. Complutense. 28040 Madrid. Spain.}
}

\author{D.Fern\'andez-Fraile}{
  address={Nuclear Theory Group,
Department of Physics, Brookhaven National Laboratory, Upton, NY-11973,
USA.}
}

\begin{abstract}
 We review  recent results on  properties of the meson gas relevant for Heavy Ion Collision and Nuclear Matter experiments, within the framework of chiral lagrangians. In particular, we describe the temperature and density evolution of the $\sigma$ and $\rho$ poles and its connection with chiral symmetry restoration, as well as the chemical nonequilibrated phase and   transport coefficients.
\end{abstract}

\maketitle


\section{Hot and dense resonances in unitarized chiral perturbation theory}

It is well established nowadays that QCD undergoes a chiral restoration transition, as confirmed by different lattice simulations \cite{Aoki:2009sc,Cheng:2009zi}. The transition is presumably a crossover for 2+1 massive light
 flavors, which means that the critical point could be different for different observables. In the lattice, the main order parameters are
  the quark condensate and the light scalar susceptibility. Nevertheless, there have been several efforts in the literature to try to find signals of chiral symmetry restorations in physical observables which could be eventually measured in Heavy Ion or Nuclear Matter experiments. The most promising signals are related to the behaviour of resonances in hot and dense matter,  such as  the $\rho-a_1$ degeneracy \cite{Rapp:1999ej,Cabrera:2009ep} or the role of the $\rho$ resonance in the dilepton spectrum \cite{Rapp:1999ej,David:2006sr}. In the latter case, a significant dropping of the $\rho$ mass would be interpreted as a signal of chiral restoration according to the Brown-Rho  scaling \cite{br} or QCD sum rules \cite{Hatsuda:1995dy} scenarios. However, this does not seem to be the case experimentally in Heavy Ion Collisions, where the alternative explanation of broadening dominance \cite{Rapp:1999ej} is equally valid to explain the data and it is actually favored in recent experiments, such as NA60 with dimuons \cite{NA60}. At RHIC energies (PHENIX), even  the broadening picture is insufficient to explain the observed excess in the region of low invariant mass \cite{phenix,Drees:2009xy}. The situation is  less clear in cold nuclear matter experiments. The E325-KEK collaboration \cite{Naruki:2005kd} has reported a measurable shift
in the masses of vector mesons compatible with theoretical predictions based on
Brown-Rho scaling  and QCD sum rules. On the
other hand,  the JLab-CLAS experiment \cite{:2007mga}  has obtained results
compatible with vanishing mass shift, as predicted by most  in-medium hadronic
many-body analysis, where broadening is the dominant effect
\cite{Urban:1998eg,Cabrera:2000dx}.

The $\sigma/f_0(600)$ meson is also a suitable candidate to study signals of chiral restoration through its in-medium behaviour, since it has the quantum numbers of the vacuum. A simple $O(4)$ model description   would suggest that the dropping of $\langle \sigma \rangle\sim \sqrt{\cond}$ driven by chiral restoration should imply a significant reduction of the $\sigma$ mass, at least in the chiral limit. Such a reduction could produce experimentally an enhancement of $\pi\pi$ scattering and cross section near the point where $M_\sigma\rightarrow 2m_\pi$ and the phase space shrinks. This is the so called threshold enhancement effect \cite{hatku85},  whose signals are observed in cold nuclear matter reactions $\pi A\rightarrow \pi\pi
A'$ \cite{Bonutti:2000bv,cb} and  $\gamma A\rightarrow \pi\pi A'$
\cite{messetal}.

Our approach is to consider the temperature and density modifications of the lightest meson resonances in the unitarized chiral lagrangian framework \cite{GomezNicola:2002tn,Dobado:2002xf,FernandezFraile:2007fv,Cabrera:2008tja}. We rely mostly on Chiral Perturbation Theory (ChPT), which provides a model-independent low-energy expansion for a given $\pi\pi$ scattering partial wave with total isospin $I$ and angular momentum $J$ as $t^{IJ}(s;T)=t^{IJ}_2(s) + t^{IJ}_4(s;T) +
\Od(p^6)$ with $s$  the center of mass energy squared and $T$ the temperature. According to the usual ChPT scheme,  $t_2$ is the tree level contribution from the lowest order lagrangian ${\cal L}_2$, while $t_4$  includes both the tree level from ${\cal L}_4$ and the one-loop diagrams from ${\cal L}_2$. The latter contain the dependence with the temperature \cite{GomezNicola:2002tn}, as well as the imaginary part demanded by unitarity, which at finite temperature reads $\im t_4=\sigma_T\vert t_2\vert^2$, where:
\begin{equation}\sigma_T (s)=\sigma_0(s)[1+2n(\sqrt{s}/2)]\label{thps}\end{equation}
is the two-pion thermal phase space, with $n(x)=[\exp(x/T)-1]^{-1}$  the Bose-Einstein distribution
function and $\sigma_0(s)=\sqrt{1-4m_\pi^2/s}$. The ChPT series cannot reproduce a resonant behaviour, which can be recovered by using a unitarization method, which amounts to find a unitary amplitude $t$ satisfying exactly $\im t=\sigma_T\vert t \vert^2$ and matching the ChPT series  when expanded at low energies or temperatures. We follow the Inverse Amplitude Method (IAM) which reads:
\begin{equation}
t^{IAM}=\frac{t_2(s)^2}{t_2(s)-t_4(s;T)+A(s;T)}
\end{equation}
where  $A(s;T)$ is a known function constructed from $t_2$ and $t_4$ ensuring that the zeros of the unitarized amplitude coincide with those
 of the perturbative one and which can be alternatively obtained using dispersion relations \cite{GomezNicola:2007qj}. It is irrelevant for poles far from the real axis but should be included in medium analysis where, as commented above, chiral symmetry may induce a vanishing imaginary part and in fact the $A$ function in that case prevents the appearance of spurious poles \cite{FernandezFraile:2007fv}.

\begin{figure}
$\begin{array}{cc}
  \includegraphics[height=.21\textheight]{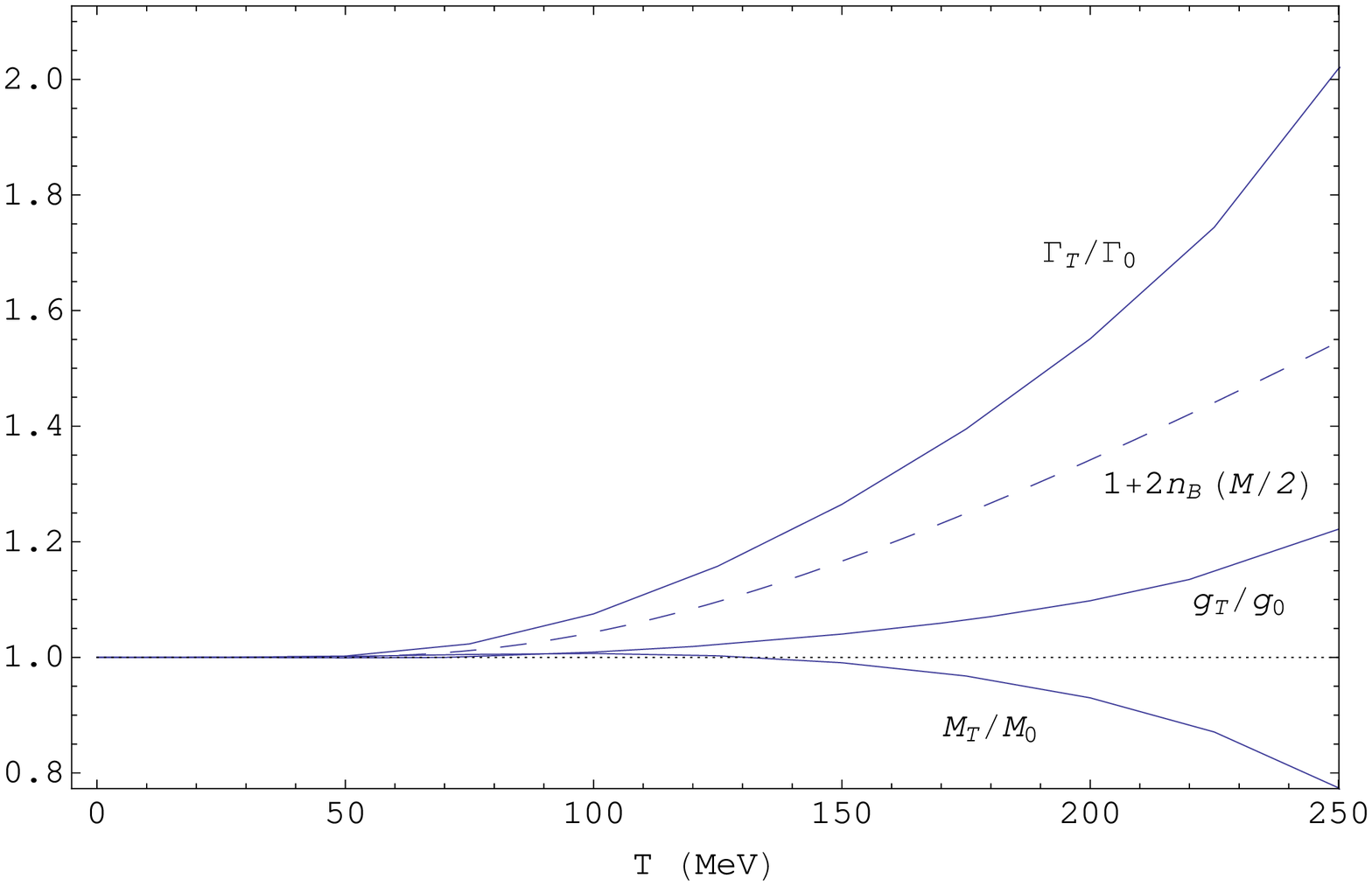}&
  \includegraphics[height=.21\textheight]{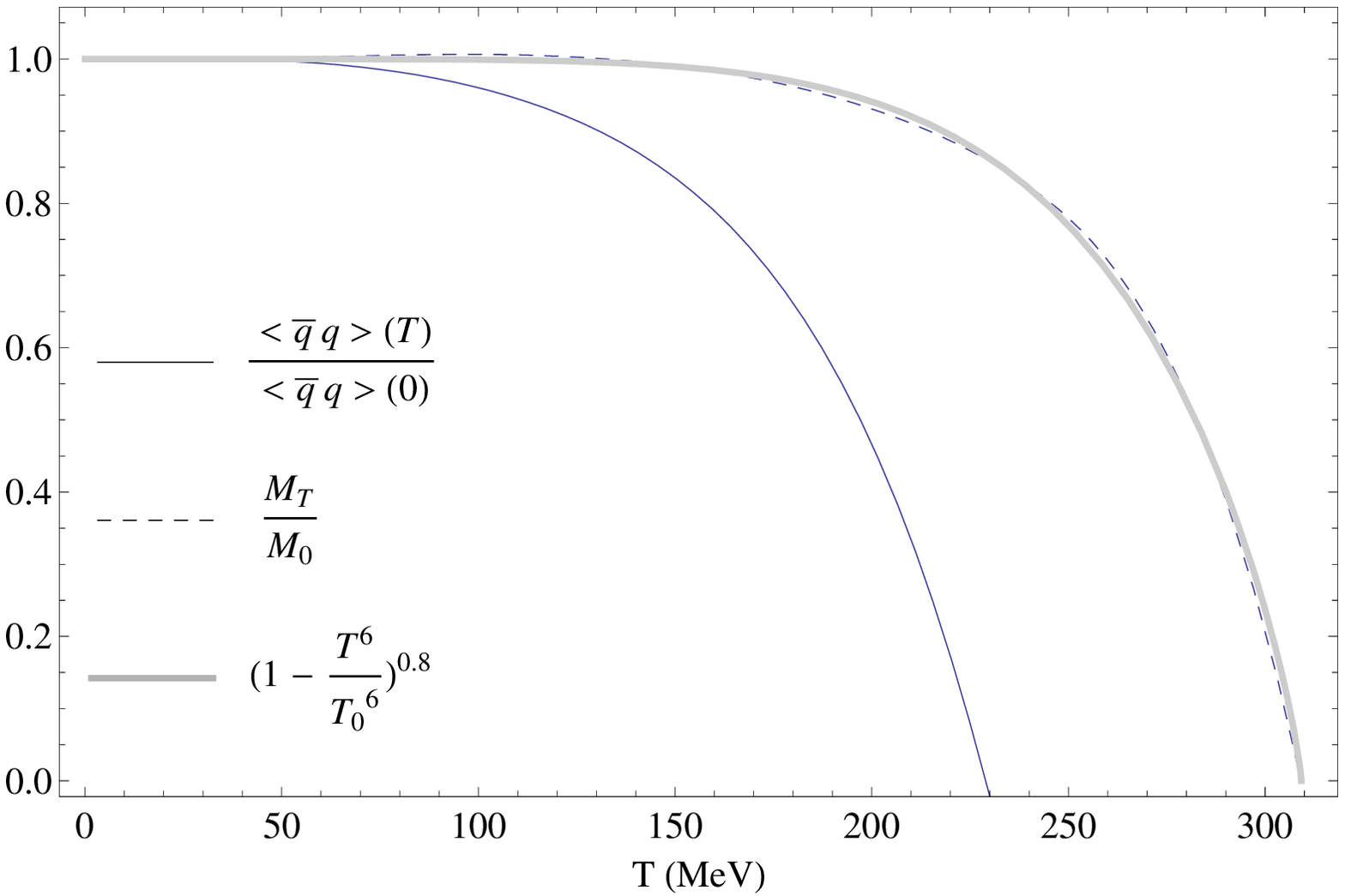}\\
  \includegraphics[height=.21\textheight]{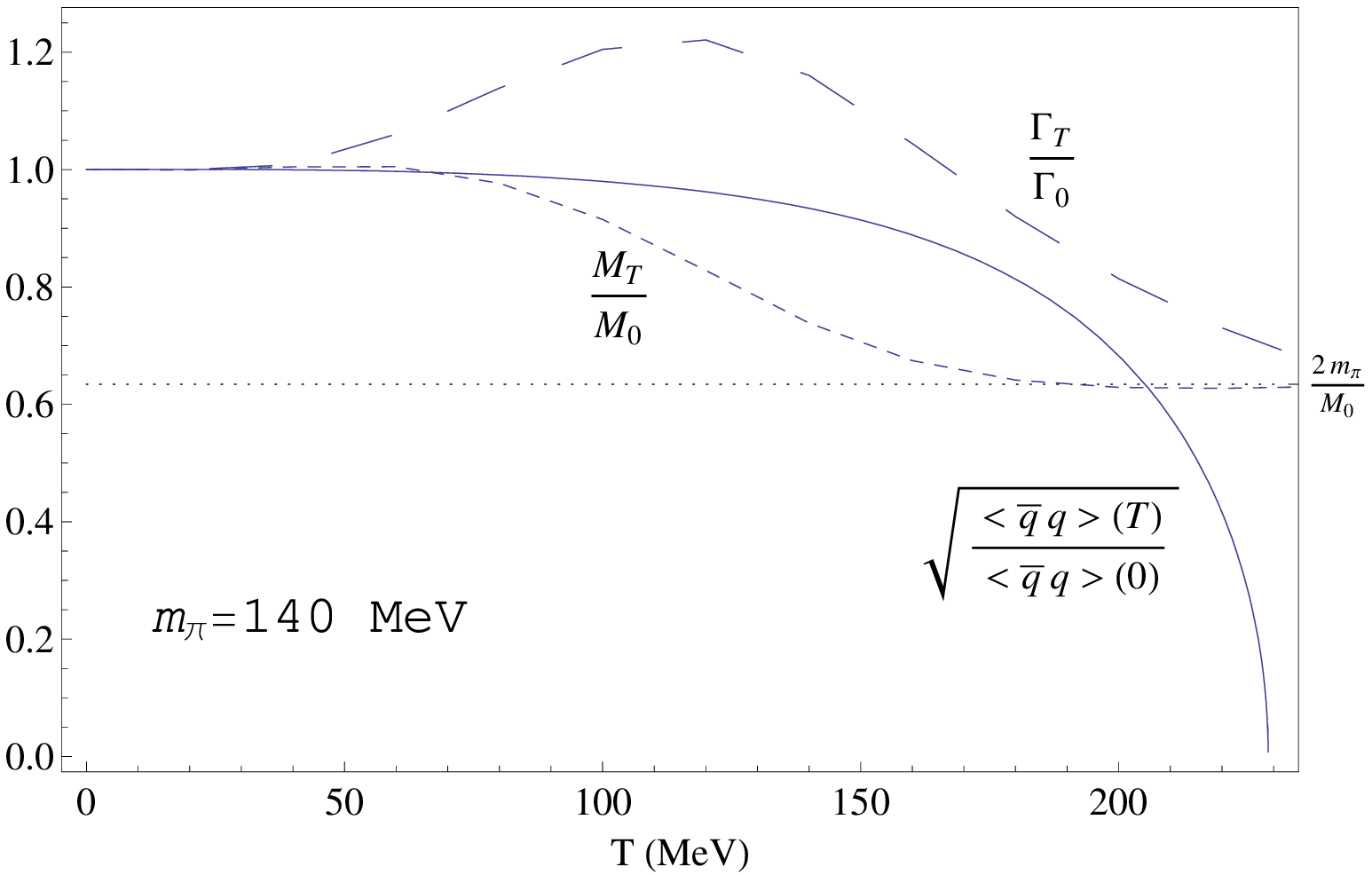}&
  \includegraphics[height=.21\textheight]{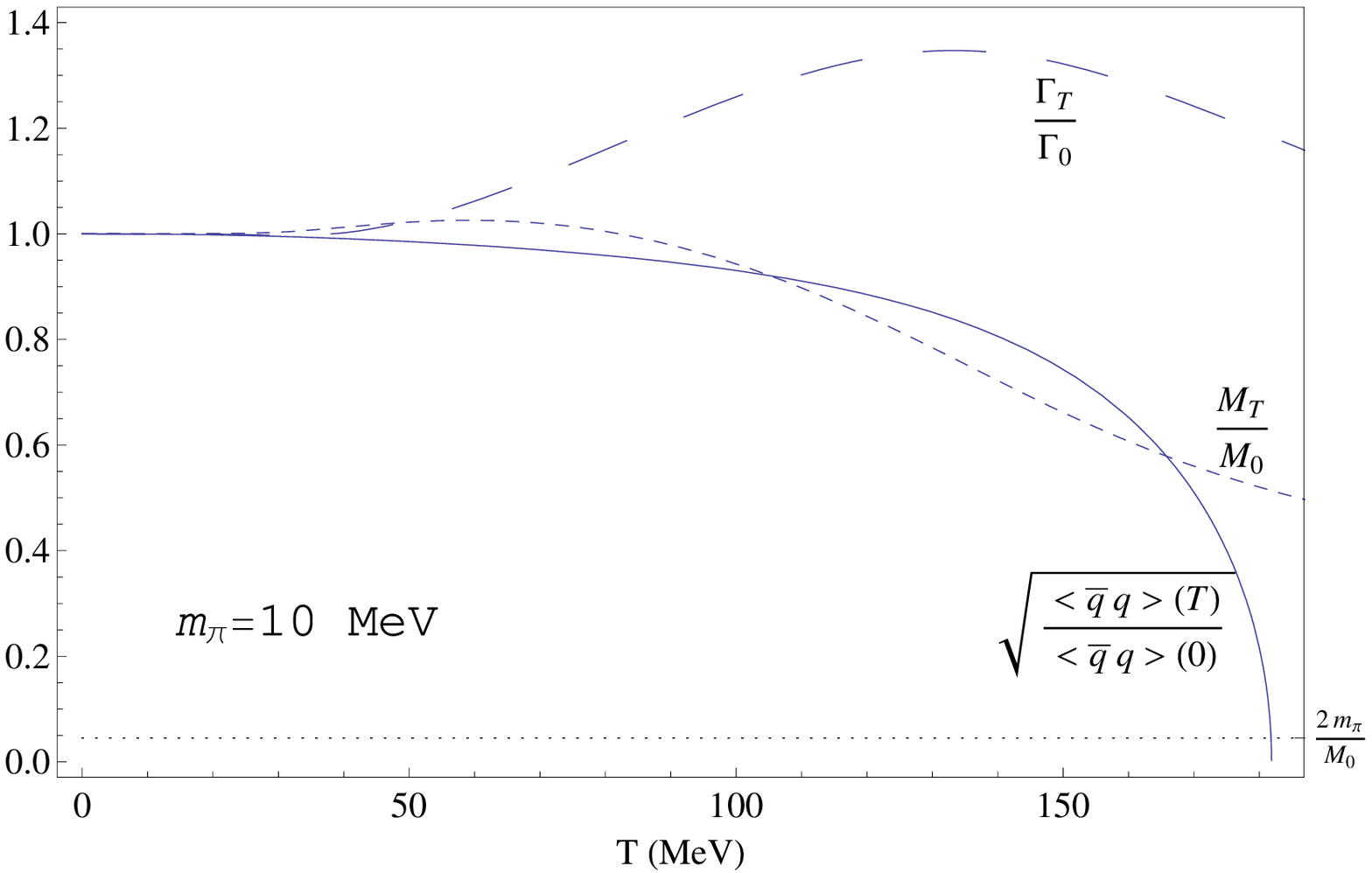}
  \end{array}$
     \caption{\label{Fig:rhosigmaT} Results for the $\rho$ (upper panels) and $\sigma$ (lower panels) thermal mass and width extracted from the IAM poles at finite temperature. }
\end{figure}

Once we have a unitarized $\pi\pi$ amplitude, the $\sigma$ and $\rho$ are generated as poles  in the second Riemann sheet dynamically, i.e., without introducing them explicitly in the lagrangian nor assuming anything about their nature. The results with the finite-$T$ IAM are showed in Fig.\ref{Fig:rhosigmaT}. The so-called low-energy constants of ${\cal L}_4$ are fixed to yield the mass and width of the $\rho$ at their physical values at $T=0$. The main conclusions of our analysis are: i) The main effect for the $\rho$ is thermal broadening, in agreement with dilepton data, the mass being reduced only slightly for temperatures below chiral restoration, ii) such broadening does not only come from phase space increasing $\sigma_T/\sigma_0$ but also from the increasing of the effective $\rho\pi\pi$ vertex, extracted from the residue of the amplitude at the pole position, iii) in the $\sigma$ channel,  mass reduction dominates, presumably driven by chiral restoration, and at some point overcomes the thermal width phase space increasing so that the imaginary part (width) reaches a maximum value and then starts decreasing, iv) Near chiral restoration, the width of the $\sigma$ remains sizable (its spectral function is not peaked around the mass as in a Breit-Wigner resonance) so that no threshold enhancement is expected in a Heavy Ion  environment, v) The mass dropping pattern in both the $\sigma$ and $\rho$ cases does not scale with the condensate as expected from Brown-Rho scaling or simple $O(4)$ model arguments for the $\sigma$, not even near the chiral limit.

\begin{figure}
  \includegraphics[height=.21\textheight]{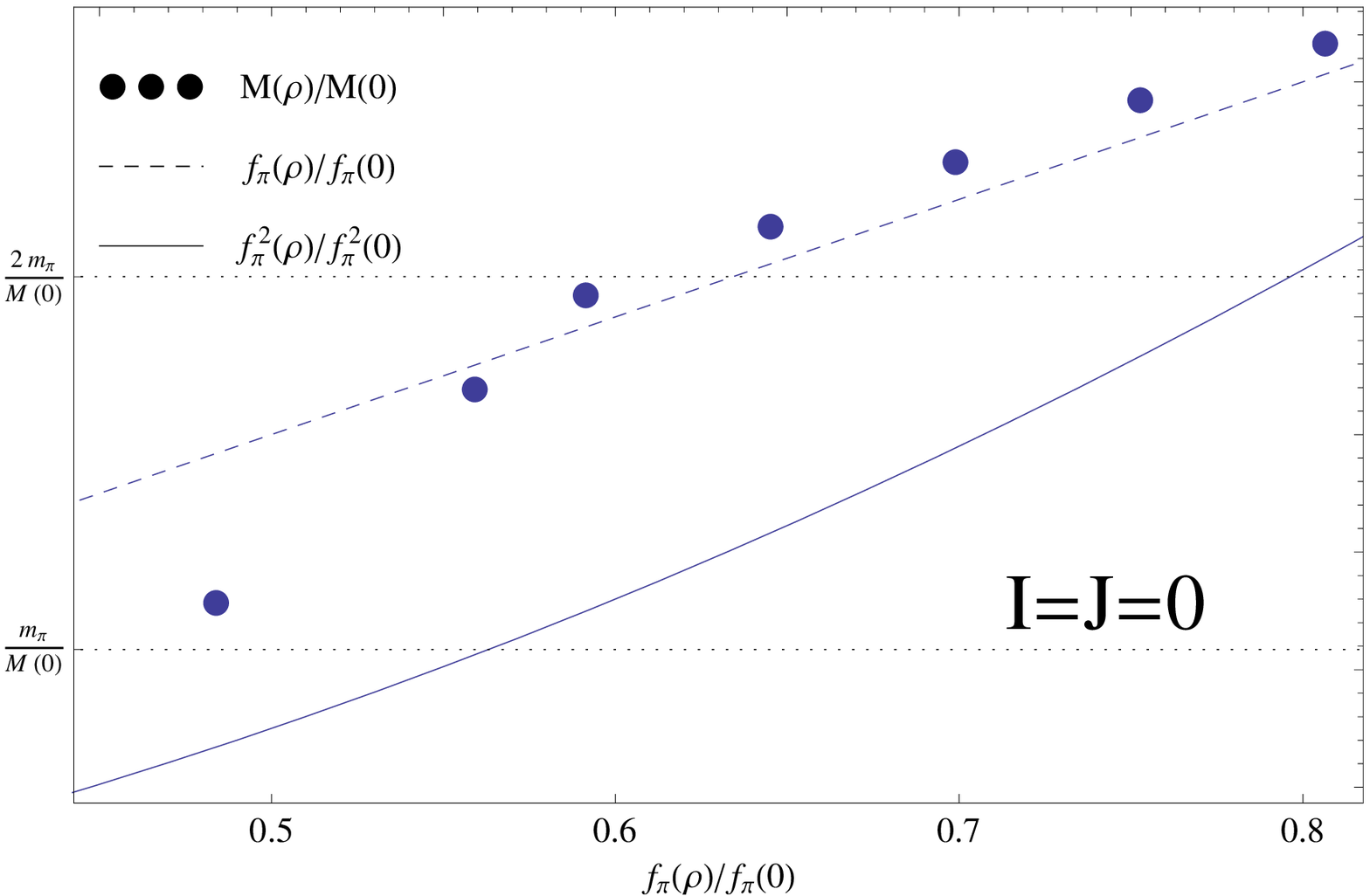}
  \includegraphics[height=.21\textheight]{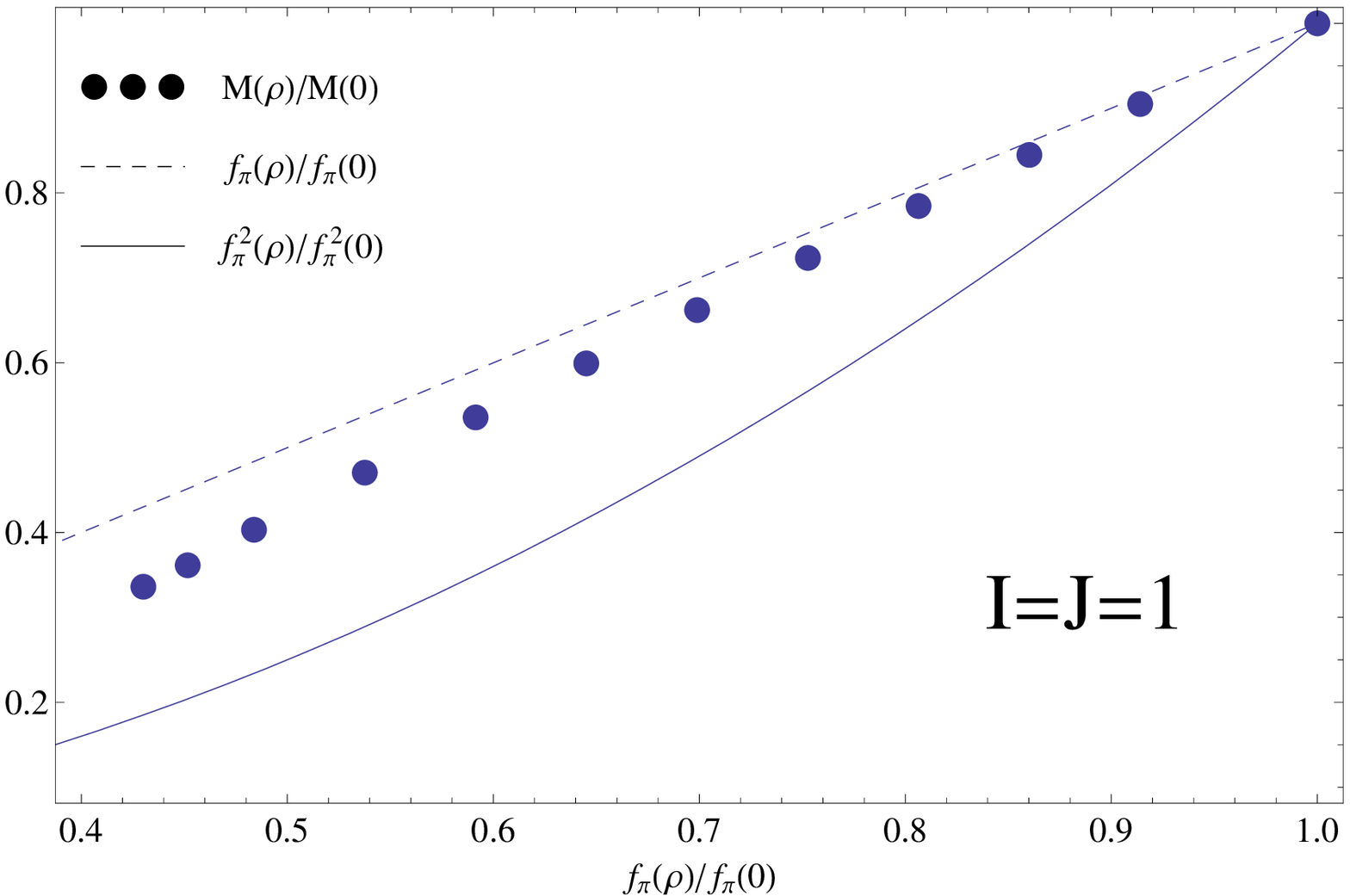}
     \caption{\label{Fig:scalingfpi} Scaling of the mass in the IAM, compared with  that of $f_\pi(\rho)$. In the $I=J=0$ channel, the masses displayed correspond to the lowest masses of the
 second-sheet poles.}
\end{figure}

The next question is how to incorporate the effect of the nuclear density $\rho$ in our approach. At $T=0$, a simple procedure \cite{FernandezFraile:2007fv} which encodes most of the relevant corrections is to scale the pion decay constant $f_\pi$ according to \cite{thowir95oller02}:
\begin{equation} \frac{f_\pi^2(\rho)}{f_\pi^2(0)}\simeq \frac{\langle \bar q
q\rangle(\rho)}{\langle \bar q q\rangle(0)}\simeq \left(1-
\frac{\sigma_{\pi N}}{m_\pi^2 f_\pi^2(0)}\rho\right)\simeq \left(
1-0.35\frac{\rho}{\rho_0}\right) \label{fpidensity} \end{equation} where $\rho$
is the nuclear density, $\sigma_{\pi N}\simeq$ 45 MeV is the
pion-nucleon sigma term and $\rho_0\simeq$ 0.17 fm$^{-3}$ is the
normal or saturation nuclear matter density. In this way we are actually examining the scaling properties of the resonances when we scale $f_\pi^2(\rho)$, or equivalently the quark condensate. Actually, in this simple approach there is no broadening source, so that only the masses  change. The results of the mass scaling in our IAM approach are showed in Fig.\ref{Fig:scalingfpi}. In the $I=J=0$ channel, the scaling corresponds to the $\sigma$ of the $O(4)$ model. When the pole approaches the real axis, threshold enhancement is clearly observed and when it reaches it a pole-doubling occurs: one of the poles remains near threshold and eventually jumps into the first Riemann sheet becoming a $\pi\pi$ bound state, while the other one (plotted in Fig.\ref{Fig:scalingfpi}) tends to degenerate with the pion as its chiral partner. Although this phenomenon takes place at rather high  $\rho\sim 2\rho_0$ for this approach to be fully trusted, it has been obtained also in \cite{patkos03}  and reveals the molecular-like character of the $f_0$ state \cite{FernandezFraile:2007fv}. The chiral degeneracy between the dynamically generated $\sigma$ and the pion for $f_\pi\rightarrow 0$ had been also noticed in \cite{Oller:2000wa}. In the $\rho$ channel, the $f_\pi$  scaling induces a linear $M_\rho$ scaling in agreement with the experimental value of \cite{Naruki:2005kd} and the theoretical predictions of \cite{br} and \cite{Hatsuda:1995dy}, which however must be taken with a pinch of salt because all the broadening effects coming from many-body interactions are neglected.

\begin{figure}
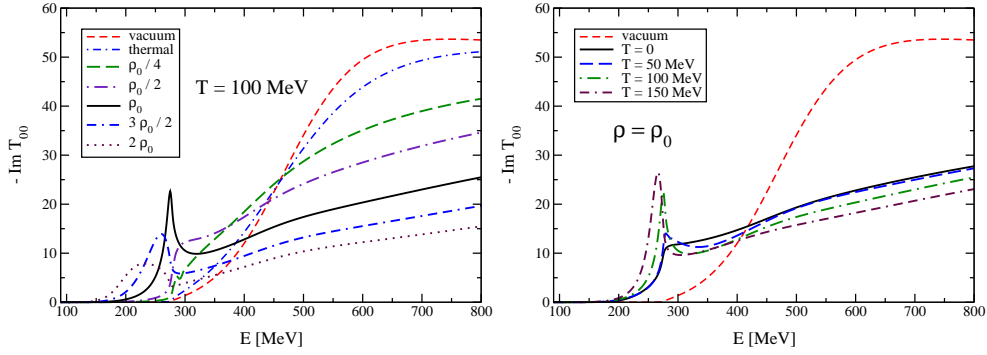

  \includegraphics[height=.21\textheight]{ImT.T100.vs.rho.eps}
  \includegraphics[height=.21\textheight]{ImT.rho0.vs.T.eps}
     \caption{\label{Fig:tempdens} Imaginary part of the $\pi\pi$ amplitude in the $I=J=0$ channel at finite
 temperature and nuclear density in the many-body BS approach.}
\end{figure}

The picture that emerges from our analysis is then that when broadening effects are ignored, chiral restoration in the $\sigma$ channel takes place more or less along the expected $O(4)$ pattern. However, the $T\neq 0$ study shows that thermal broadening distorts the picture of a $\bar q q$-like narrow state and, despite experiencing a large mass reduction, the state is still wide near the transition. This is confirmed by the detailed many-body calculation performed in \cite{FernandezFraile:2007fv} for the $\sigma$ channel, including both nuclear density and temperature effects.  The unitarization scheme used is the Bethe-Salpeter one, more suitable for nuclear many-body analysis, and takes into account the pion self-energy  in the nuclear medium plus all the other relevant diagrams compatible with chiral symmetry at the same order in density. The results in Figure \ref{Fig:tempdens} for the imaginary part of the amplitude show some distinctive features of this full approach: the abrupt change with respect to the vacuum or thermal case, originated by physical in-medium excitations such as particle-hole or $\Delta$-hole, which produce also strength below threshold, the sizable threshold enhancement, even though the pole is still rather away from the axis like in the thermal case, and finally the amplification of the threshold strength by the combined effect of temperature and density, which in fact accelerates the migration of the pole towards threshold by sudden mass decrease.
\section{Pions out of chemical equilibrium}

In the hadron gas formed in a Relativistic Heavy Ion Collision, it is phenomenologically well justified to consider a phase
 where thermal equilibrium still prevails but inelastic collisions involving particle number change are negligible, so that chemical equilibrium is
  lost. For instance, according to the estimates in \cite{Song:1996ik}, at $T=150\ \mathrm{MeV}$ the relaxation time of elastic $\pi\pi$ collision is $\tau_{el}\sim 2 \mathrm{fm}/c$, whereas that of the process $\pi\pi\leftrightarrow \pi\pi\pi\pi$ is $\tau_{in}\sim 200\ \mathrm{fm}/c$, the hadronic phase lifetime being about $10 \ \mathrm{fm}/c$. In other words, there is a temperature window between the chemical freeze-out temperature $T_{chem}\sim$ 180 MeV, below which approximately only elastic collisions remain, and the thermal freeze-out one $T_{ther}\sim$ 100-120 MeV of hadron decoupling, where the chemical potential associated to pion number conservation $\mu_\pi$ is not zero \cite{Bebie:1991ij,Hung:1997du,Kolb:2002ve}. Actually, in this phase $\mu_\pi$ is a function of $T$ so that $\mu_\pi(T\rightarrow T_{chem})=0$. On the other hand, different phenomenological fits of particle yields and ratios at SPS and RHIC energies predict $\mu_\pi(T_{ther})\sim$ 70-100 MeV \cite{Hung:1997du,Kolb:2002ve}.

\begin{figure}
  \includegraphics[height=.21\textheight]{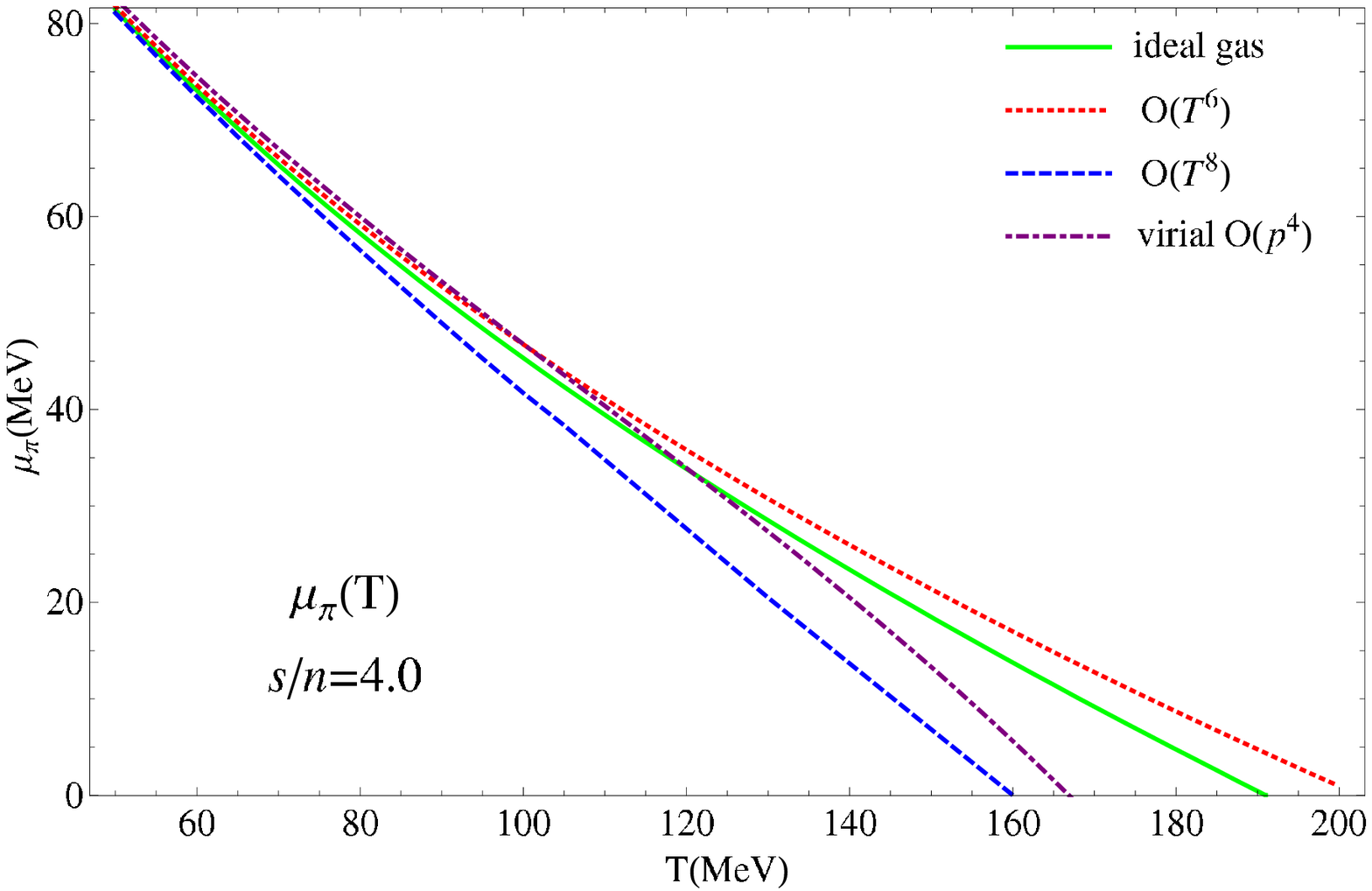}
  \includegraphics[height=.21\textheight]{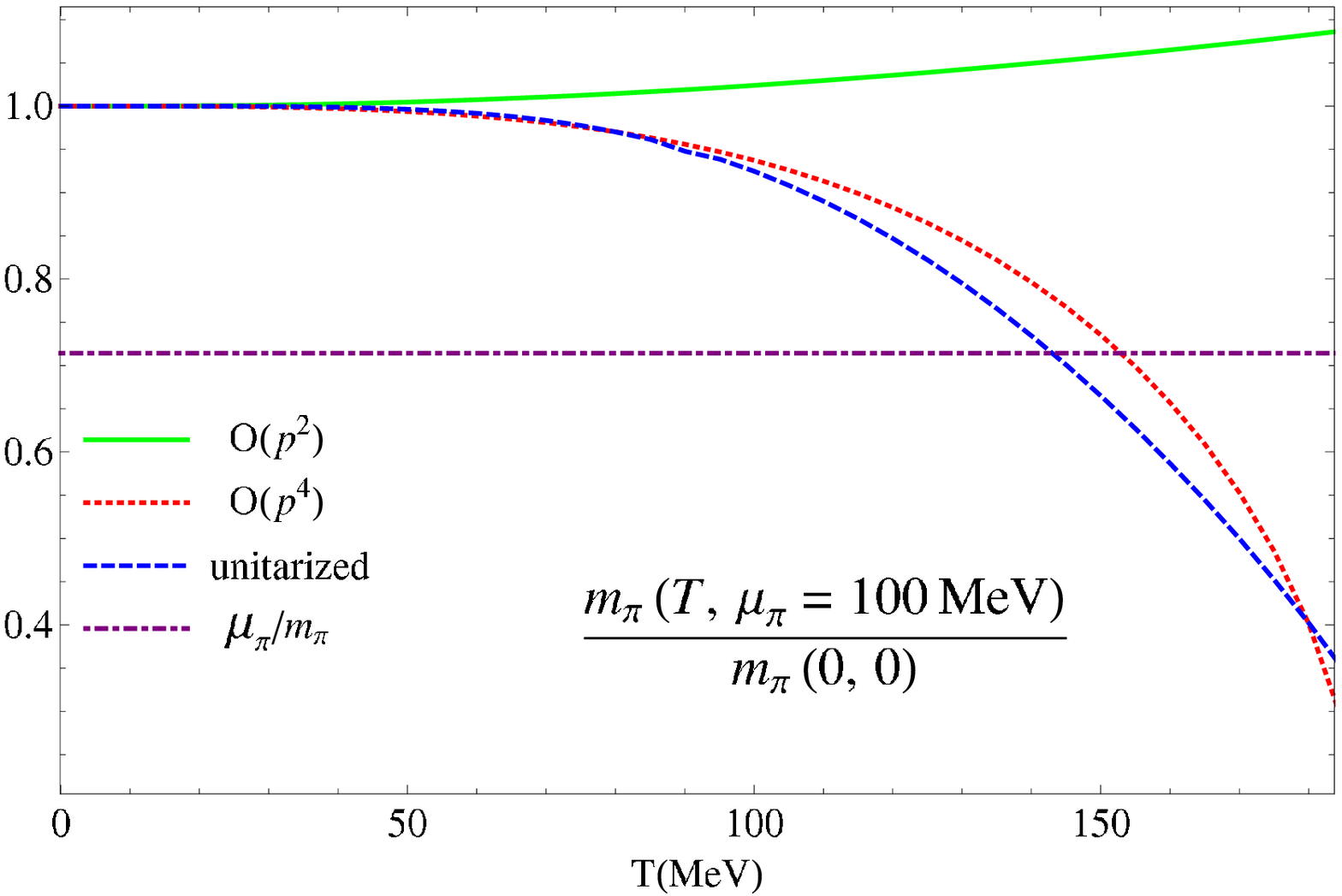}
      \caption{\label{Fig:muT} Left: the function  $\mu_\pi(T)$ with the isentropic condition for different approximations. Right: the pion mass dependence with $T$ and $\mu_\pi$.}
\end{figure}

  In a recent work \cite{FernandezFraile:2009kt}, we have considered a ChPT description of this phase in the pion gas, within a Quantum Field Theory treatment of chemical nonequilibrium for boson fields. A crucial point is that the total pion number is only approximately conserved, unlike for instance the electric charge or the third isospin component. This makes the usual path-integral description in terms of field states inconvenient for this case. We have used instead an holomorphic representation, which allows to develop the formalism in terms of creation and annihilation operators, in terms of which one can easily express the free particle number.  Another technical complication is that there are no KMS-like periodicity conditions, which prevents the imaginary-time thermal formalism to be used. As in other nonequilibrium formulations, the appropriate choice is a real-time contour in complex time. This implies the doubling of the pion degrees of freedom, so that the propagators have a two-dimensional matrix structure. One of the two types of fields is unphysical and only appears in internal lines. After a detailed analysis, we obtain  the generating functional, which can be read from the usual one at $\mu_\pi=0$ with the following replacements for the free propagators and for the free partition function:
   \begin{eqnarray}
D_{11}(p_0,E_p)&=&\frac{i}{p_0^2-E_p^2+i\epsilon}+2\pi\delta(p_0^2-E_p^2)
n(\vert
p_0 \vert-\mu_\pi)
\nonumber\\
 D_{22}(p_0,E_p)&=&
\frac{-i}{p_0^2-E_p^2+i\epsilon}+2\pi\delta(p_0^2-E_p^2)n(\vert
p_0 \vert-\mu_\pi)\nonumber\\
 D_{12}(p_0,E_p)
&=&
2\pi\delta(p_0^2-E_p^2)\left[\theta(-p_0)+n(\vert
p_0 \vert-\mu_\pi)\right]=D_{21}(-p_0,E_p)\nonumber\\
\log
Z_\beta^0&=&-V\int\frac{\mathrm{d}^3\vec{p}}{(2\pi)^3}\
\left[\frac{\beta
E_p}{2}+\log\left(1-\mathrm{e}^{-\beta(E_p-\mu)}\right)\right]
\label{rtpropmom}
\end{eqnarray}
with $V$ the system volume and $E_p^2=\vert\vec{p}\vert^2+m_\pi^2$. With the above ingredients, we can calculate the different thermodynamical variables of the pion gas in this regime, by
 evaluating the corresponding closed diagrams, whose topology is the same as the $\mu_\pi=0$ case \cite{Gerber:89}. We
  have extended our analysis up to $\Od(T^8)$. At that order, particle-changing processes can be seen to become already important, but the perturbative scheme  remains consistent since an additional physical condition has to be imposed in order to regulate the plasma dynamical evolution  from thermal to chemical equilibrium, namely the $\mu_\pi(T)$ function. A simple way to model this is to demand that the ratio of entropy density to pion particle number density $s/n$ remains constant \cite{Bebie:1991ij}. This isentropic condition is nothing but the combination of entropy conservation without dissipation and approximate particle number conservation in this phase. The crucial point is that this ratio is a decreasing function of both $T$ and $\mu_\pi$, so that as the systems cools down, the chemical potential grows in order to keep it constant. The precise value of $s/n$ can be fixed either by the chemical freeze-out temperature at which $\mu_\pi(T_{chem})=0$ or by the value $\mu_\pi(T_{ther})$ at thermal freeze-out.  We show in Figure \ref{Fig:muT} our results for different chiral orders, comparing also with the virial expansion approach followed in \cite{Dobado:1998tv}. We see that one of the consequences of including the pion interactions is to lower the value of $T_{chem}$ by about 25 MeV with respect to the ideal gas.

Our diagrammatic approach allows also to calculate the pion self-energy corrections in $T$ and $\mu_\pi$. The detailed analysis and results are given in \cite{FernandezFraile:2009kt}. To leading order in the pion density (dilute gas regime) the self-energy obeys a Luscher-like relation:
\begin{eqnarray}
m_\pi^2(T,\mu_\pi)-m_\pi^2(0,0)=-\int
\frac{d^3{\vec{p}}}{(2\pi)^3}\frac{n(E_p-\mu_\pi)}{2E_p}
\re T^f_{\pi\pi} (s=(E_p+m_\pi)^2-\vert\vec{p}\vert^2)+\Od(n^2)\nonumber\\
\Gamma_p(T,\mu_\pi)=\frac{1}{2E_p}\int \frac{d^3\vec{k}}{(2\pi)^3} \frac{n(E_k-\mu_\pi)}{2E_k}
 \im T^f_{\pi\pi}\left[s=(E_p+E_k)^2-\vert \vec{p}+\vec{k}\vert^2\right]+\Od(n^2)
\label{gamma}
\end{eqnarray}
where $T^f_{\pi\pi}(s)$ is the forward $\pi\pi$ scattering amplitude. The $\Od(p^4)$ and unitarized (IAM) amplitudes produce a decreasing pion mass with both $T$ and $\mu_\pi$. Actually, at a given temperature, the mass can reach the chemical potential, as showed in Figure \ref{Fig:muT}. This implies the interesting possibility that pion Bose-Einstein condensation could be reached dynamically, i.e., driven by interactions. The corresponding $\mu^{BE}(T)$ where this condition is met is above but not far from the isentropic curves for realistic chemical freeze-out conditions. On the other hand, from the pion width $\Gamma_p$, we can estimate the elastic mean collision time $\tau_{el}$ and  the thermal freeze-out temperature where $\tau_{el}(T_{ther})\sim$ 10 fm/c. Taking into account the $\mu_\pi$ dependence of $\tau_{el}$ through the isentropic $\mu_\pi(T)$ yields a reduction $\Delta T_{ther} \sim$ -20 MeV with respect to neglecting chemical nonequilibrium.

\section{Transport coefficients}

Transport coefficients measure the linear response of the system to a deviation from equilibrium induced by an external source. Their knowledge is essential to describe many phenomenological aspects regarding dissipation effects in Heavy Ion Collisions.
Recently, we have developed an extensive programme for the calculation of transport coefficients within diagrammatic ChPT, with direct application to the mesonic phase of the plasma expansion \cite{FernandezFraile:2005ka,FernandezFraile:2008vu,FernandezFraile:2009mi}. The diagrammatic treatment is technically nontrivial because transport coefficients involve the zero external frequency and momentum limit of retarded correlators. This implies the appearance of the so called pinching poles, proportional to $1/\Gamma$ with $\Gamma$ the collisional width of the particles in the medium. This generates nonperturbative contributions, since $\Gamma$ is proportional to the collision amplitude (see e.g. eq.(\ref{gamma})) and therefore perturbatively small. Thus, the usual chiral power counting of ChPT has to modified to account properly for these contributions \cite{FernandezFraile:2005ka}.
\begin{figure}
$\begin{array}{c}
\includegraphics[height=.22\textheight]{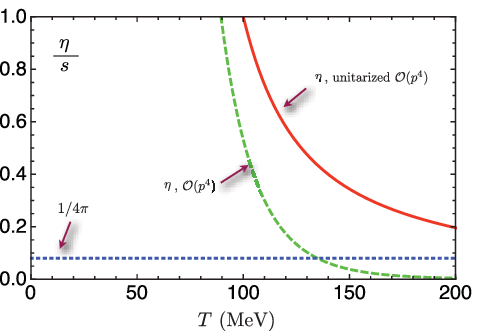}\\
\includegraphics[height=.21\textheight]{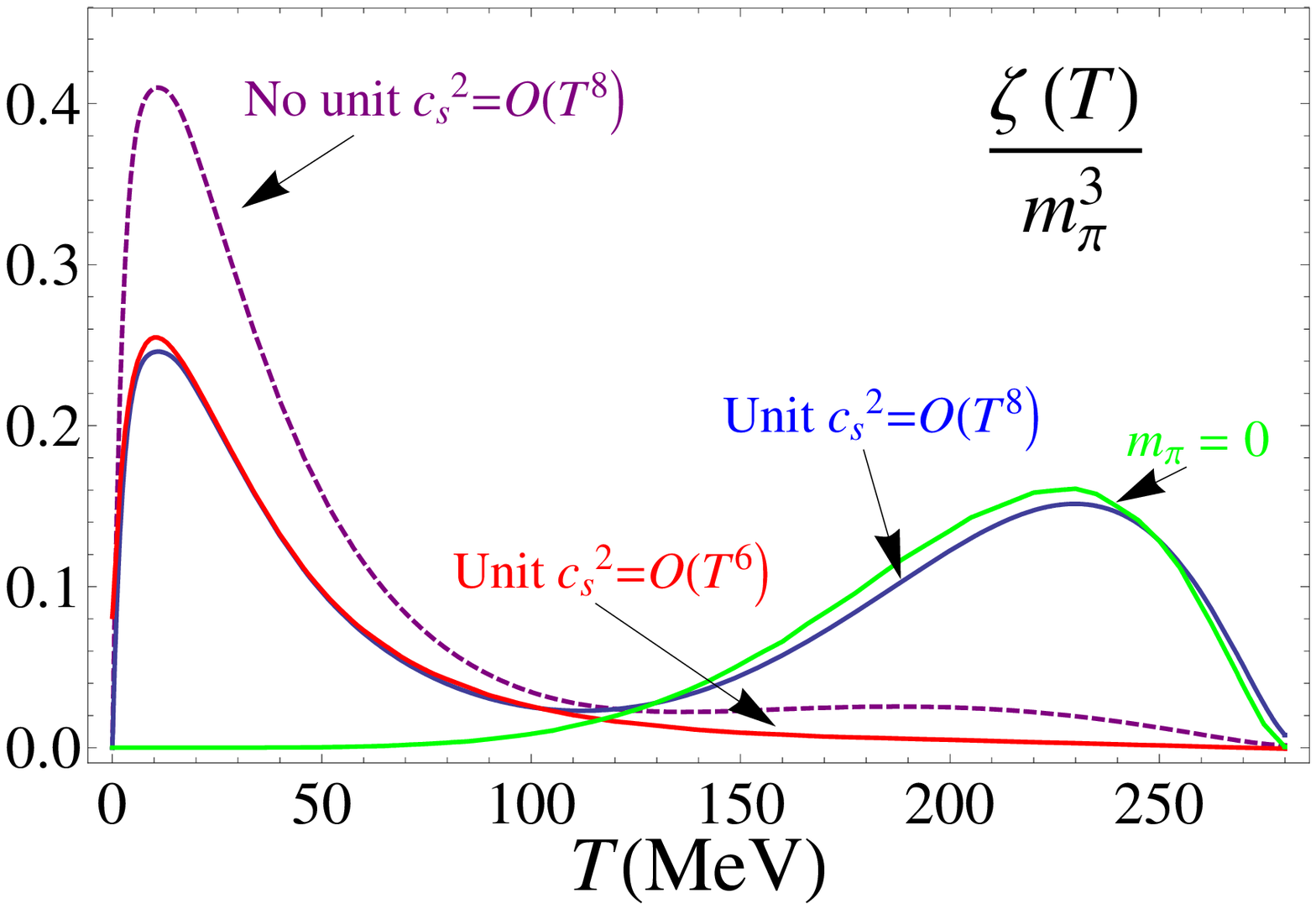}
  \includegraphics[height=.21\textheight]{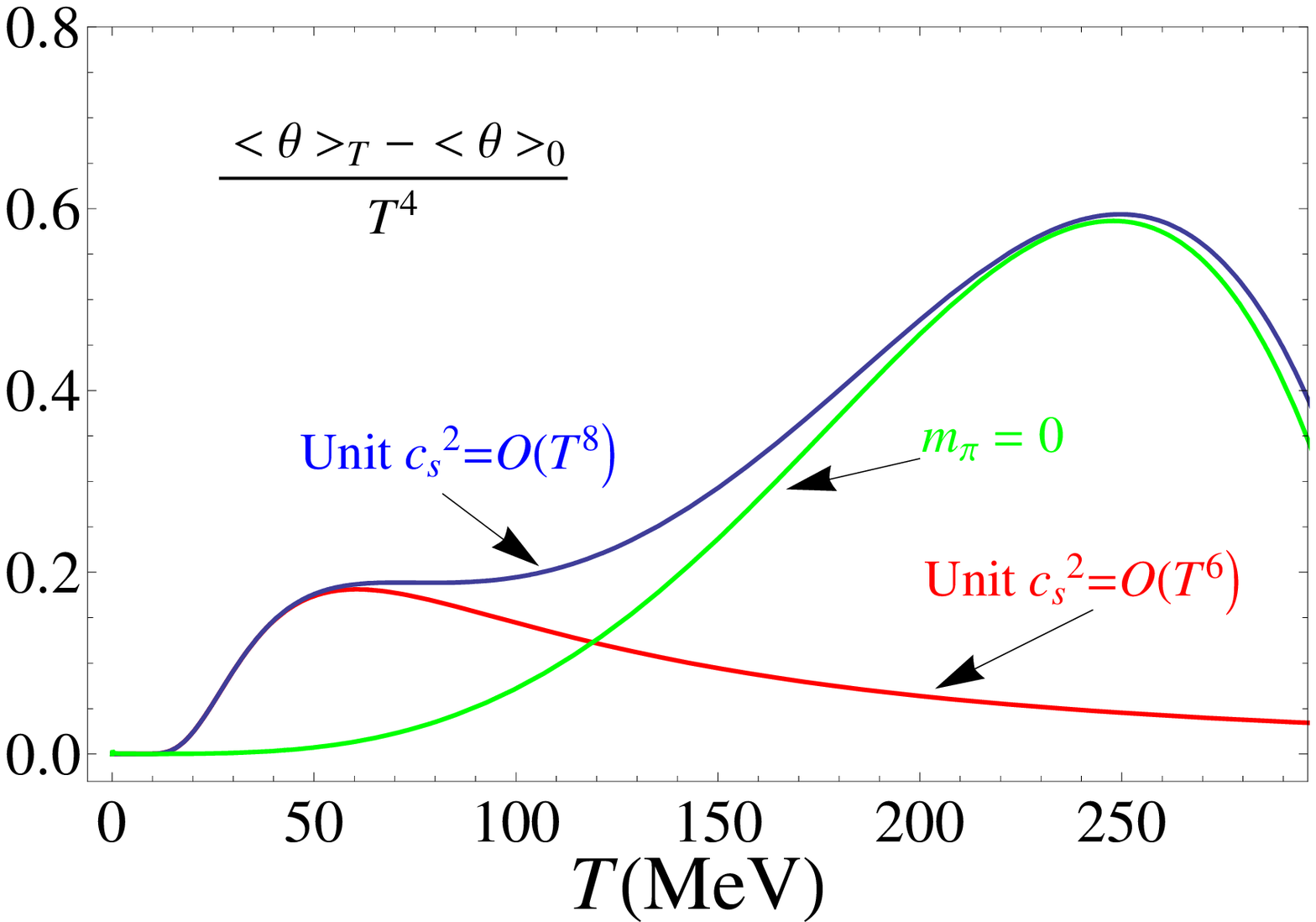}
\end{array}$
\caption{\label{Fig:visco} Viscosity coefficients for the pion gas in ChPT. Up: shear viscosity. Down: Bulk viscosity and  trace anomaly.}
\end{figure}

The viscosity coefficients are particularly interesting for phenomenology. The shear viscosity to entropy density ratio $\eta/s$ can be determined through  elliptic flow analysis, the present RHIC data pointing to a small value  $\eta/s < 0.5$ \cite{Dusling:2007gi}, an almost perfect fluid. Our ChPT result is greatly influenced by a correct unitarized description of the scattering amplitudes entering  the pion width. In fact, without unitarization, at $\Od(p^4)$, $\eta/s$ is a monotonically decreasing function of $T$, violating the so called KSS bound $\eta/s>1/4\pi$ \cite{Kovtun:2004de}, as showed in Figure \ref{Fig:visco}. The unitarized result satisfies that bound,  gives phenomenologically compatible values at the relevant temperatures and is  compatible with the existence of a minimum for this ratio near the transition \cite{Dobado:2008vt}. The case of the bulk viscosity $\zeta$ is also important because it is directly sensitive to variations of conformal symmetry. Thus,
 it should have a similar behaviour \cite{Karsch:07} to the trace anomaly:
\begin{equation}
\langle\theta\rangle_T\equiv\langle T^\mu_\mu\rangle_T=
T^5\frac{\mathrm{d}}{\mathrm{d}T}\left(\frac{P}{T^4}\right)
\end{equation}
where $T^\mu_\nu$ is the energy-momentum tensor and $P$ the thermodynamic pressure. The trace anomaly has a maximum at the QCD transition, originated mostly from the anomalous glue contribution to the beta function and observed in the lattice \cite{Cheng:2009zi}. Therefore, significantly large  values for $\zeta$ could be observed near the critical point. Our results  in Figure
\ref{Fig:visco} show a clear correlation between these quantities, both developing a two-peak structure. The first maximum comes from the explicit conformal breaking due to the quark mass and thus it disappears in the chiral limit. Actually, in that temperature range  $\delta\langle\theta\rangle_T\sim 2m_q \delta\cond_T$ \cite{FernandezFraile:2008vu}. The second maximum is the critical one and
 is almost unchanged in the chiral limit,  not being related to the quark condensate or to the quark mass, but to deconfinement effects
  encoded in the gluon condensate anomalous conformal breaking.
  We see again that incorporating properly unitarization  is crucial to describe correctly the high temperature region.
  It is also essential  to consider the speed of sound squared $c_s^2$ to $\Od(T^8)$ in order to capture its minimum at the transition
  arising from  anomalous conformal breaking. Our analysis does not rely on assumptions on the behaviour of the $\theta\theta$ spectral function, as it is the case of the original proposal in \cite{Karsch:07}, which has been recently corrected \cite{Romatschke:09}. It is therefore unclear from general considerations how the correlation between the trace anomaly and the bulk viscosity should actually hold. In fact, it does not hold in simple models mimicking QCD \cite{FernandezFraile:2010gu}. We believe then that our pion gas description can
  be useful to understand this and other issues regarding the behaviour of transport coefficients below the critical temperature.
\begin{theacknowledgments}
  Work partially supported by the Spanish
research contracts FPA2008-00592,  FIS2008-01323, UCM-BSCH GR58/08 910309.
  \end{theacknowledgments}
\bibliographystyle{aipproc}   

\begin{thebibliography}{99}
\bibitem{Aoki:2009sc}
  Y.~Aoki {\it et al.},
  JHEP {\bf 0906}, 088 (2009).
\bibitem{Cheng:2009zi}
  M.~Cheng {\it et al.},
  Phys.\ Rev.\  D {\bf 81}, 054504 (2010).
\bibitem{Rapp:1999ej}
  R.~Rapp and J.~Wambach,
  Adv.\ Nucl.\ Phys.\  \textbf{25}, (2000) 1.
  \bibitem{Cabrera:2009ep}
  D.~Cabrera, D.~Jido, R.~Rapp and L.~Roca,
  Prog.\ Theor.\ Phys.\  {\bf 123}, 719 (2010).
\bibitem{David:2006sr}
  G.~David, R.~Rapp and Z.~Xu,
  Phys.\ Rept.\  {\bf 462}, 176 (2008).
\bibitem{br} G.~E.~Brown and M.~Rho,
  Phys.\ Rev.\ Lett.\  \textbf{66}, (1991) 2720;
  Phys.\ Rept.\  {\bf 363}, 85 (2002).
\bibitem{Hatsuda:1995dy}
  T.~Hatsuda, S.~H.~Lee and H.~Shiomi,
  Phys.\ Rev.\  C {\bf 52} (1995) 3364.
\bibitem{NA60} R.~Arnaldi {\it et al.}  [NA60 Collaboration],
  Phys.\ Rev.\ Lett.\  \textbf{96}, (2006) 162302.
\bibitem{phenix}
  A.~Adare {\it et al.}  [PHENIX Collaboration],
  Phys.\ Lett.\  B {\bf 670}, 313 (2009).
\bibitem{Drees:2009xy}
  A.~Drees,
  Nucl.\ Phys.\  A {\bf 830}, 435C (2009).
\bibitem{Naruki:2005kd}
  M.~Naruki {\it et al.},
  Phys.\ Rev.\ Lett.\  {\bf 96} (2006) 092301.
\bibitem{:2007mga}
  R.~Nasseripour {\it et al.}  [CLAS Collaboration],
  Phys.\ Rev.\ Lett.\  \textbf{99}, (2007) 262302.
\bibitem{Urban:1998eg}
  M.~Urban, M.~Buballa, R.~Rapp and J.~Wambach,
  Nucl.\ Phys.\  A {\bf 641} (1998) 433.
\bibitem{Cabrera:2000dx}
  D.~Cabrera, E.~Oset and M.~J.~Vicente Vacas,
  Nucl.\ Phys.\  A \textbf{705}, (2002) 90.
\bibitem{hatku85}
  T.~Hatsuda and T.~Kunihiro,
  Phys.\ Rev.\ Lett.\  \textbf{55}, (1985) 158.
\bibitem{Bonutti:2000bv}
  F.~Bonutti {\it et al.}  [CHAOS collaboration],
  Nucl.\ Phys.\  A \textbf{677}, (2000) 213.
\bibitem{cb} A.~Starostin {\it et al.}  [Crystal Ball Collaboration],
  Phys.\ Rev.\ Lett.\  \textbf{85}, (2000) 5539.
\bibitem{messetal} J.~G.~Messchendorp {\it et al.},
  Phys.\ Rev.\ Lett.\  \textbf{89}, (2002) 222302.
\bibitem{GomezNicola:2002tn}
  A.~Gomez Nicola, F.~J.~Llanes-Estrada and J.~R.~Pelaez,
  Phys.\ Lett.\  B {\bf 550}, 55 (2002).
\bibitem{Dobado:2002xf}
  A.~Dobado, A.~Gomez Nicola, F.~J.~Llanes-Estrada and J.~R.~Pelaez,
  Phys.\ Rev.\  C {\bf 66}, 055201 (2002).
\bibitem{FernandezFraile:2007fv}
  D.~Fernandez-Fraile, A.~Gomez Nicola and E.~T.~Herruzo,
  Phys.\ Rev.\  D {\bf 76}, 085020 (2007).
\bibitem{Cabrera:2008tja}
  D.~Cabrera, D.~Fernandez-Fraile and A.~Gomez ~Nicola,
  Eur.\ Phys.\ J.\  C {\bf 61}, 879 (2009).
\bibitem{GomezNicola:2007qj}
  A.~Gomez Nicola, J.~R.~Pelaez and G.~Rios,
  Phys.\ Rev.\  D {\bf 77} (2008) 056006.
\bibitem{thowir95oller02} V.~Thorsson and A.~Wirzba,
   Nucl.\ Phys.\ \textbf{A589}, (1995) 633.  U.~G.~Meissner, J.~A.~Oller and A.~Wirzba,
  Annals.\ Phys.\  \textbf{297}, (2002) 27.
\bibitem{patkos03} A.~Patkos, Z.~Szep and P.~Szepfalusy,
  Phys.\ Rev.\  \textbf{D68}, (2003) 047701.
\bibitem{Oller:2000wa}
  J.~A.~Oller,
  arXiv:hep-ph/0007349.
\bibitem{Song:1996ik}
  C.~Song and V.~Koch,
  Phys.\ Rev.\  C {\bf 55}, 3026 (1997).
\bibitem{Bebie:1991ij}
  H.~Bebie, P.~Gerber, J.~L.~Goity and H.~Leutwyler,
  Nucl.\ Phys.\  B {\bf 378}, 95 (1992).
\bibitem{Hung:1997du}
  C.~M.~Hung and E.~V.~Shuryak,
  Phys.\ Rev.\  C {\bf 57}, 1891 (1998).
\bibitem{Kolb:2002ve}
  P.~F.~Kolb and R.~Rapp,
  Phys.\ Rev.\  C {\bf 67}, 044903 (2003).
\bibitem{FernandezFraile:2009kt}
  D.~Fernandez-Fraile and A.~Gomez Nicola,
  Phys.\ Rev.\  D {\bf 80}, 056003 (2009).
\bibitem{Gerber:89}
  P.Gerber and H.Leutwyler, Nucl.\ Phys.\ B{\bf 321}, 387 (1989).
\bibitem{Dobado:1998tv}
  A.~Dobado and J.~R.~Pelaez,
  Phys.\ Rev.\  D {\bf 59}, 034004 (1999).
\bibitem{FernandezFraile:2005ka}
  D.~Fernandez-Fraile and A.~Gomez Nicola,
  Phys.\ Rev.\  D {\bf 73}, 045025 (2006).
\bibitem{FernandezFraile:2008vu}
  D.~Fernandez-Fraile and A.~G.~Nicola,
  Phys.\ Rev.\ Lett.\  {\bf 102}, 121601 (2009).
\bibitem{FernandezFraile:2009mi}
  D.~Fernandez-Fraile and A.~Gomez Nicola,
  Eur.\ Phys.\ J.\  C {\bf 62}, 37 (2009).
\bibitem{Dusling:2007gi}
  K.~Dusling and D.~Teaney,
  Phys.\ Rev.\  C {\bf 77}, 034905 (2008).
\bibitem{Kovtun:2004de}
  P.~Kovtun, D.~T.~Son and A.~O.~Starinets,
  Phys.\ Rev.\ Lett.\  {\bf 94}, 111601 (2005).
\bibitem{Dobado:2008vt}
  A.~Dobado, F.~J.~Llanes-Estrada and J.~M.~Torres-Rincon,
  Phys.\ Rev.\  D {\bf 79}, 014002 (2009).
\bibitem{Karsch:07} F.Karsch, D.Kharzeev and K.Tuchin, Phys.Lett.B{\bf 663}, 217 (2008).
\bibitem{Romatschke:09} P. Romatschke and D. T. Son, Phys. Rev. D, 2009, {\bf 80}: 065021.
\bibitem{FernandezFraile:2010gu}
  D.~Fernandez-Fraile,
  arXiv:1009.2741 [hep-ph].
\end{thebibliography}


\end{document}